\documentclass[running heads]{svmult}

\usepackage{makeidx}   
\usepackage{graphicx}  
\usepackage{subeqnar}  
\usepackage{multicol}  
\usepackage{physprbb}  
\makeindex             

\newcommand{\greeksym}[1]{{\usefont{U}{psy}{m}{n}#1}}
\newcommand{\umu}{\mbox{\greeksym{m}}}

\newcommand{\gr}{$\gamma$-ray\,}
\newcommand{\grs}{$\gamma$-rays\,}
\begin{document}
\title*{Astrophysics with High Energy Gamma Rays}
\toctitle{Astrophysics with High Energy Gamma Rays}
%
%
\titlerunning{High Energy Gamma Rays}
%
\author {Heinrich J. V\"olk}
\tocauthor{H.J.~V\"olk}
\authorrunning{H.J. V\"olk}
%
%
\institute{Max-Planck-Institut f\"ur Kernphysik, 
Saupfercheckweg 1,\\
D-69117 Heidelberg, Germany}

\maketitle              

\begin{abstract}
Recent results, the present status and the perspectives of high energy
gamma-ray astronomy are described. Since the satellite observations by the
Compton Gamma Ray Observatory and its precursor missions have been
reviewed extensively, emphasis is on the results from the ground-based
gamma-ray
telescopes. They concern the physics of Pulsar Nebulae, Supernova Remnants
in their assumed role as the Galactic sources of Cosmic Rays, Jets from
Active Galactic Nuclei, and the Extragalactic Background radiation field
due to stars and dust in galaxies. Since the gamma-ray emission is
nonthermal, this kind of astronomy deals with the pervasive high-energy
nonequilibrium states in the Universe. The present build-up of larger and
more sensitive instruments, both on the ground and in space, gives
fascinating prospects also for
observational cosmology and astroparticle physics. Through realistically
possible further observational developments at high mountain altitudes a
rapid extension of the field is to be expected.
\end{abstract}

\section{Introduction}
There is a general consensus that the main energy sources for high energy
\gr emission are extreme objects in the Universe with high energy
turnover. This belief stems from the observation of young Pulsars,
Supernova Remnants (SNRs) and Active Galactic Nuclei (AGNs), and we shall
discuss some of these detections here.

We know that high energy \grs are abundantly produced in collisions of
charged particles that have been accelerated in collective processes. They
involve ionized systems (i.e. collisionless plasmas), large scale mass
motions, and electromagnetic fields. The nonthermal processes are
important because they are part of the major energy dissipation processes
like shock waves that arise in explosive events, at the breaking of
supersonic flows in the form of winds and jets from galaxies, and during
mergers of galaxies and clusters. Another important source of particle
acceleration should be the dissipative angular momentum transport in
magnetized accretion flows near compact objects. Therefore we can
plausible assume that a major fraction of the random kinetic energy in the
Universe is nonthermal, with many of the particles at ultrarelativistic
energies. And we can see them in gamma rays.

High energy \grs might also be due to rare decays of ultra-heavy particles
or arise from annihilations of the lightest supersymmetric particles that
are widely believed to make up the nonbaryonic Dark Matter. Such indirect
identifications with the aid of \gr observations are goals of
astroparticle physics, even though no effects have been found until now.
All \gr detectors include nevertheless the fundamental issue of Dark
Matter search as part of their observation programs.

Physically speaking, the most interesting \gr features are {\it localized}
emissions. They should immediately portray the generating energetic
processes and thus give us new insights into the astrophysics of the
sources. Localized emissions from deep gravitational potential wells like
the center of our Galaxy also appear as the most promising indicators for
accumulations of weakly interacting massive cold Dark Matter particles
(WIMPs).

From the point of view of Cosmic Ray (CR) physics \gr astronomy is an
indirect form of energetic particle detection. But it is a crucial one
since even ultrarelativistic charged particles are strongly deflected from
straight line orbits by the interstellar and intergalactic magnetic
fields. The direct detection of the charged particles will therefore not
help us to find their sources even if they happen to reach the Earth. A
possible exception are the CRs with the highest energies $\sim 10^{20}$~eV
(see the paper by A.A. Watson in these Proceedings). At the much lower
energies $\leq 10^{15}$~eV, where energetic particle distributions
typically contain almost all their energy \emph{density}, only neutral
secondary collision products like \grs or neutrinos point back to the
sources. Due to their different production modes and their vastly
different interaction strength with matter they should give complementary
results. Intensive R\&D efforts are presently made to develop neutrino
astronomy, using large volumes of polar ice and ocean water, as discussed
by F. Halzen in these Proceedings. I shall concentrate here on high energy
\gr astronomy.

Since space born \gr astronomy has been reviewed extensively in the past
(e.g. \cite{sch}), I will put more emphasis the recent results from
ground-based telescopes and their impact on major physics questions. The
astronomical objects that have been successfully studied are as diverse as
Pulsar Nebulae, Supernova Remnants, AGN jets, and the diffuse
Extragalactic background radiation field in the Optical/Infrared
wavelength range. At the end I will indicate the perspectives of the
overall field for the future.

\section{High Energy \gr Detectors}
High energy \grs are measured with pair production detectors. They have
been used on satellites up to energies of the order of 10 GeV and on the
ground above a threshold of about 200 GeV, with special detector
arrangements reaching energies as low as 50 GeV.

\begin{figure}
\includegraphics[width=.5\textwidth]{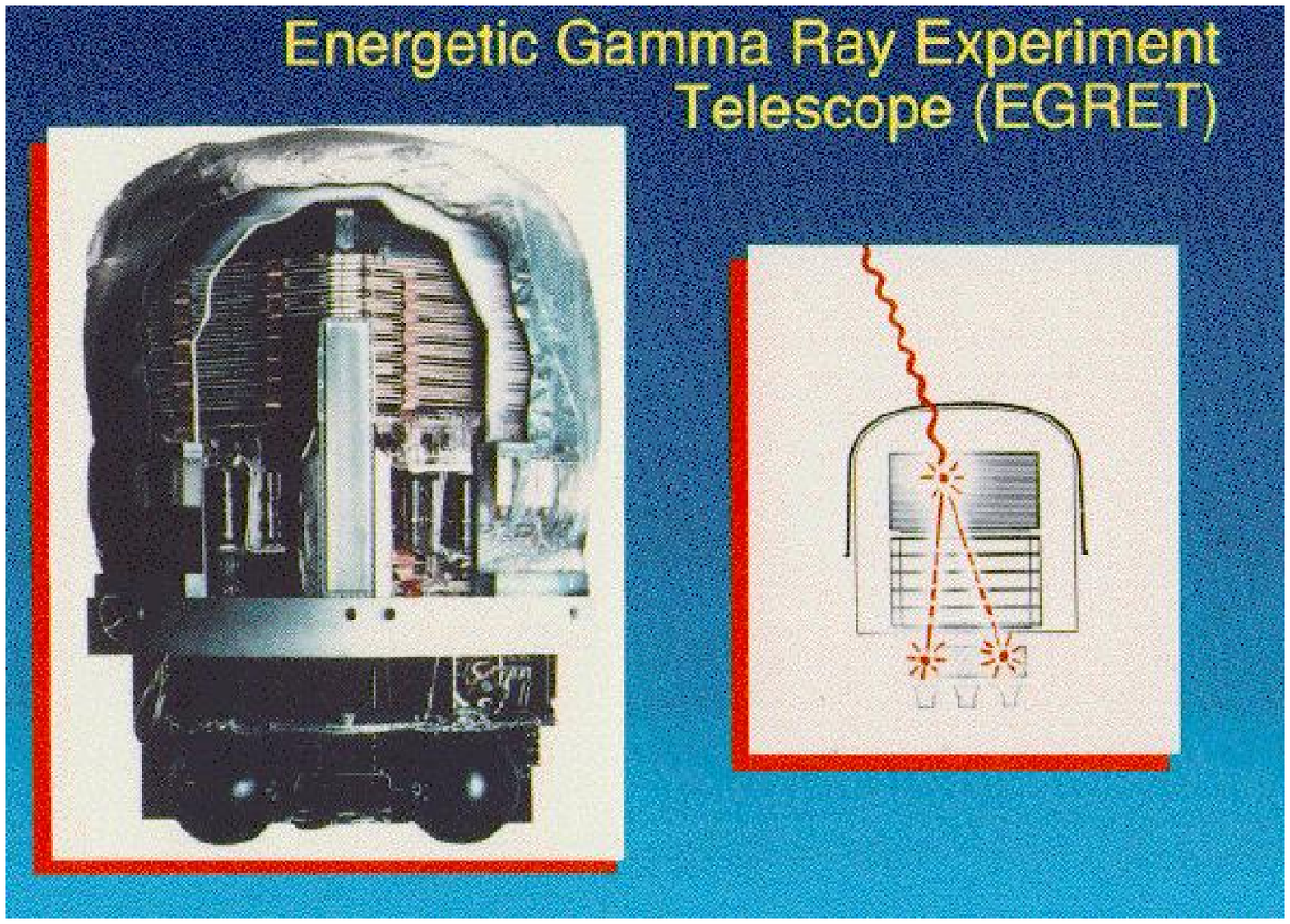}
\hspace*{\fill}
\includegraphics[width=.45\textwidth]{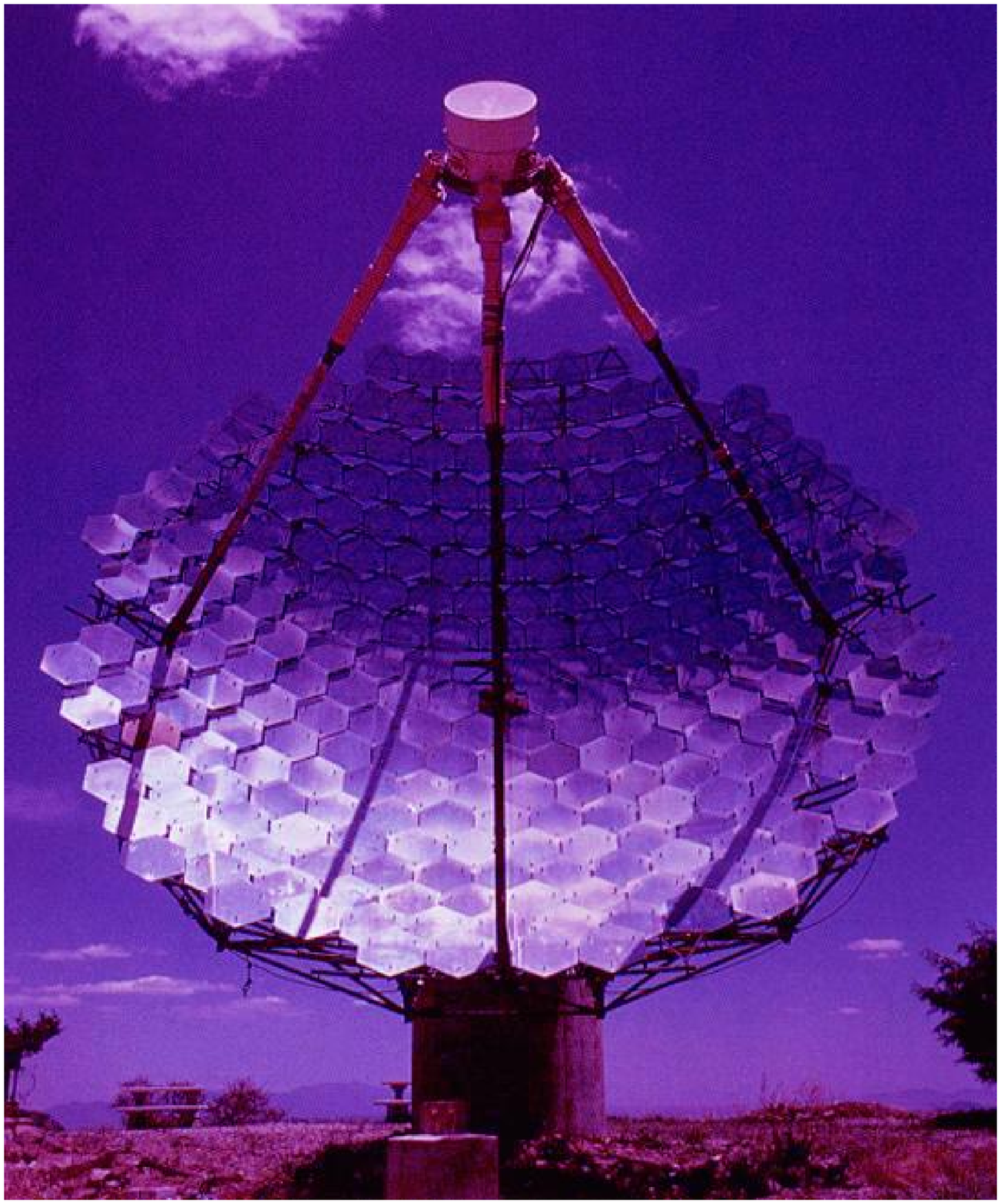}
\vspace*{2mm}

\caption[]{(Left) The EGRET instrument on the Compton Gamma Ray Observatory
(CGRO). Left: overall instrument with anti-coincidence shield. Right:
schematic of conversion of the incoming primary \gr into an $\E ^{+}\E 
^{-}$~pair that is tracked in a spark chamber, with energy measurement in
the calorimeter on the bottom}
\label{eps1}  
\end{figure}

A typical space instrument is shown in Fig.~\ref{eps1}.  To reject the
dominant flux of charged CR particles, the detector is covered by an
anti-coincidence shield. Following the NASA telescope SAS-II, ESA's Cos B
was a remarkably long-lived mission, to be finally succeeded by the
EGRET-instrument on CGRO. These detectors had a large field of view $\sim
1$~sr and a small effective area $< 1~\mathrm{m}^2$, for energies
$30~\mathrm{MeV} \leq E \leq 30$~GeV (e.g. \cite{sch}). Since the
termination of CGRO no satellite experiment is operating in this energy
range.

%
\begin{figure}
\vspace*{-2mm}
\caption[]{(Right) The 10\,m Whipple telescope in Arizona, of the now VERITAS
collaboration. The Alt/Azimuth reflector consists of tessellated glass
mirrors that focus the atmospheric Cherenkov light onto a camera whose
pixels are fast phototubes}
\label{eps2}
\end{figure} 

At \gr energies above about 5 GeV, the atmospheric shower containing many
pairs can be used on the ground to detect the associated Cherenkov light.
Today the standard instruments are imaging optical telescopes
(Fig.~\ref{eps2}).
Thus the atmosphere itself is part of the detector. The dominant
background of showers from nuclear CRs can be suppressed by analysis of
the shower images, separating the broad hadronic showers from the
concentrated electromagnetic \gr - showers. Actually this is best achieved
with a stereoscopic array of Cherenkov telescopes, as pioneered by the
HEGRA telescope system on La
Palma\,\footnote{\tt http://www.mpi-hd.mpg.de/hfm/CT/CT.html.}.
Presently, four major instruments of this type are operating:
Whipple/VERITAS (Arizona)\,\footnote{\tt http://veritas.sao.arizona.edu/veritas/technical-details.shtml.},
CANGAROO (Australia)\,\footnote{\tt http://icrhp9.icrr.u-tokyo.ac.jp/.}, 
CAT (French Pyrenees)\,\footnote{\tt http://lpnp90.in2p3.fr/~cat/index.html.}, 
and HEGRA. 

In contrast to the satellite detectors they have a small field of view of
a few degrees and a low duty cycle $\sim 10$~\% due to the restriction to
clear and moonless nights. However the effective telescope area $\sim
\mathrm{few} \times 10^4 \mathrm{m}^2$ is extremely large, the energy
resolution achieved is $\sim 15$~\%, and the angular resolution with $\sim
0.1^{\circ}$ per event is an order of magnitude better than the satellite
telescopes flown up to now. Also non-imaging detector systems, using large
mirror areas from solar power plants, are being developed to observe at
lower thresholds $\sim 20$~GeV, like CELESTE\,\footnote{\tt http://wwwcenbg/extra/Astroparticule/celeste/index.html.}, 
STACEE\,\footnote{\tt http://www.astro.ucla.edu/~stacee/.}, 
or GRAAL\,\footnote{\tt http://rplaga.tripod.com/almeria/.}, 
but we do not have space to discuss them here.
\vspace{3mm}

\noindent
\textbf{Space and/or Ground?} 
\vspace{1mm}

\noindent
As a result of these technological
developments the two types of instruments are not only complementary in
their energy range but also in their instrumental capabilities. It is
therefore no surprise that next generation instruments are developed both
for the ground as well as for space. The costs differ by more than an
order of magnitude.

\section{Gamma-Ray Astrophysics Results at High Energies}

\subsection*{Pulsar Nebulae of Rotating Neutron
Stars and their Magnetic Fields} 

Pulsed GeV \grs have been detected from eight young Pulsars by EGRET.
This emission is usually attributed to radiation from electron - photon
cascades in the Pulsar magnetosphere (e.g. \cite{rud}), even though
recently an alternative radiation mechanism has been proposed due to
magnetic reconnection in the winds of Pulsars with streams of alternating
magnetic polarity \cite{kir}.

\begin{figure}[t]
\begin{center}
\includegraphics[width=\textwidth]{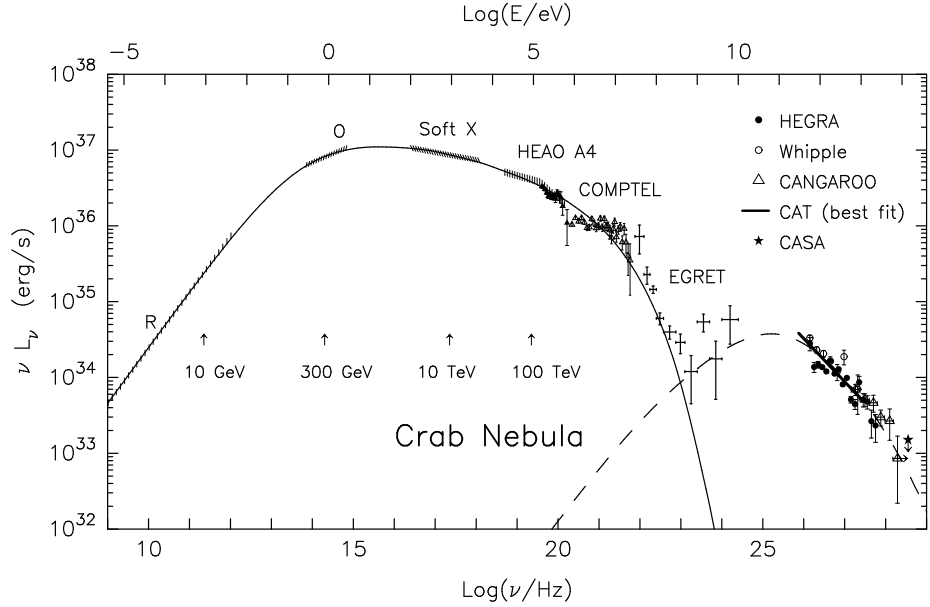}
\end{center}
\caption[]{Spectral energy density $\nu\,L_{\nu}$ of the Crab Nebula's   
emission. The high intensity synchrotron emission reaches into the \gr
range as observed by several satellite instruments. For this specific
young object it exceeds by far the flux in the second hump, presumably due
to inverse Compton radiation, as observed with several Cherenkov telescopes
(after \cite{ato})} 
\label{eps3}
\end{figure}

Unpulsed soft X-ray synchrotron emission from quite a number of extended
nebulae around Pulsars has been detected with the ROSAT and ASCA
telescopes (e.g \cite{bec}), \cite{kaw}) at rather high luminosities:
$L_{\mathrm{X}} \sim 10^{-3}~L_{\mathrm{spin-down}}$. Except for the Crab 
Nebula and PSR B1706 - 44 the implied Inverse
Compton (IC) TeV \gr emission by the same electrons in the Cosmic
Microwave Background (CMB) is only marginally measurable from most
sources at present. Yet the coming generation of \gr telescopes should be
able to detect a significant number of these objects -- and therefore the
average B-field -- because the radiant luminosities should be comparable,
$L(E_{\mathrm{IC}}/L(E_{\mathrm{syn}}) = U_{\mathrm{ph}}/U_{\mathrm{B}}
\leq 10^{-1}$ for interstellar-type magnetic fields $B \sim 10^{-5}
\mathrm{G} = B_{-5}$ which might be typical for many sources \cite{ah97};
here $U_{\mathrm{ph}}$ and $U_{\mathrm{B}}$ denote the energy densities
of the photon field and the magnetic field, respectively.
%
%
For the Crab Pulsar, and therefore presumably for other Pulsars as well,
the Nebula is thought to arise through an ultrarelativistic wind of $\E
^{+}\E ^{-}$~pairs, dissipating at a termination shock and accelerating
particles that populate a slowly expanding hot and partially nonthermal
bubble \cite{ree}, \cite{ken}, \cite{aro}. The spectrum of the emitted
radiation has been modeled phenomenologically by e.g. \cite{dej} and
\cite{ato}. For the Crab Nebula almost the entire energy loss of the
rotating neutron star is emitted by synchrotron radiation from the
Nebula, that means, in a nonthermal form (Fig.~\ref{eps3})\footnote{There
is in fact no sign for synchrotron emission from a `thermal' shocked wind
component with a characteristic energy scale expected in UV/soft X-rays,
in possible contrast to the Vela Pulsar (S. Bogovalov, private
communication). Using an argument from diffusive shock acceleration
theory we speculate that in the extremely luminous Crab Nebula the
nonlinear backreaction of the accelerated particles on the flow has
smoothed the shock completely, so that the `thermal' component is only
adiabatically compressed wind, remaining essentially `cold' (e.g.
\cite{dru}).}. This picture is of course quite simplified as the recent
highly resolved X-ray observations of the Crab with the Chandra telescope
({\tt http://chandra.harvard.edu/photo/0052/index.html}) show. But the basic
physics ingredients of the Rees \& Gunn model seem remarkably robust.
%
%

In the very young Crab Nebula 
%
%
$L(E_{\mathrm{IC}}/L(E_{\mathrm{syn}}) \sim 10^{-3}$ which implies a high
average magnetic field strength $B \simeq 3 \times 10^{-4}$~Gauss. Thus,
despite the fact that it is the strongest steady source of nonthermal
radiation in the sky -- in particular also in TeV \grs -- it is a
comparatively inefficient IC emitter due to its enormous intrinsic field
$B \gg B_{-5}$. Even though magnetic field strengths of this magnitude
are expected from equipartition arguments for this object, and had been
inferred from the spectral steepening of the synchrotron emission in the
far infrared wavelength range \cite{mar}, the independent measurements of
the synchrotron as well as the IC emission, put such inferences on a
solid experimental basis.

For the second case, PSR B1706 - 44, the steady IC flux
%
%
is about 1/5 of the Crab flux at TeV energies. However, this does not
mean a huge synchrotron luminosity by analogy. Rather we have
$L(E_{IC})/L(E_{syn}) \sim 10$, according to \cite{kif}, \cite{cha}.
Thus, either the B-field in the Nebula is on average extremely small, as
in an old extended bubble, or the particles IC scatter outside the Nebula
where the field is probably indeed very low. Large scale morphological
information may be needed to interpret the measurement \cite{ah97}.
\vspace{3mm}

\noindent
\textbf{What can we expect from future \gr measurements of a large
sample for an improved understanding of Pulsar Nebulae?}
\vspace{1mm}

\noindent
If PSR B1706 - 44 turned out to be the more typical case than the Crab
Nebula, then we might have a modified picture: In analogy to the lack of
equilibrium of nonrelativistic wind bubbles around massive stars (e.g.
\cite{bre}), the Pulsar Nebula would likely be dynamically unstable
allowing the bubble to cool, not in the form of synchrotron cooling but
rather by escape of the relativistic particle component into the
surrounding diffuse Supernova Remnant. The spatially integrated {\it
total} nonthermal emission should still be comparable to the spin-down
luminosity, with the changing magnetic field now playing primarily the
role of a wind and particle accelerator and not that of a cooling agent.

The possible appearance of a shocked thermal component of the nebula is
expected to exhibit characteristic synchrotron and IC signatures that
would directly allow a measurement of the Lorentz factor of the Pulsar
Wind. The relative strength of the thermal component may in addition be a
indicator for the degree of nonlinearity of the acceleration process.

\subsection*{Shell Type Supernova Remnants\\ 
and the Origin of Galactic Cosmic Rays}
If we disregard the possible compact remnant, the physics interest in
diffuse, shell-type Supernova Remnants (SNRs) is readily enumerated: as an
ensemble the SNRs lead to the largest mechanical energy input into the
Interstellar Medium of galaxies. The strong blast wave from the explosion,
sweeping up the circumstellar medium, suggests efficient diffusive shock
acceleration of charged nuclei to a power law source spectrum roughly
$\propto E^{-2}$. Maximum energies should possibly reach $10^{15}~eV$ and
the elemental ratios should very roughly correspond to cosmic abundances.
In short, SNRs are suspected to be the sources of the Galactic CRs.
Essentially by default.

Even SNRs fulfill the enormous energy requirement
for the replenishment of the Cosmic Rays in the Galaxy of $\sim
10^{40}$~erg/yr not by a large margin; at least 10\% of the
entire mechanical energy released in the event must on average be
converted into nonthermal energy of relativistic nuclei. 

The only experimental test of this widespread belief consists in direct
multi-wavelength observations of SNRs. Due to the inferred hard source
spectra, with about equal energy per decade, the best test uses very high
energy \grs and/or neutrinos since all background radiations fall off more
steeply with particle energy. Up to now the required sensitivity can only
be approached by \gr experiments and this prospect has been one of the
driving forces behind the development of high energy \gr astronomy.
Notwithstanding these goals, all \gr detectors operating up to now have
been only marginally sensitive in the face of existing flux estimates
\cite{dav}, \cite{nat}.

Also CR electrons are detected on the top of the atmosphere, at a 1 percent
flux level in comparison with CR nuclei. And they are equally assumed to be
accelerated in SNRs.


Although electrons contribute only to a negligible degree to the CR energy
density, a high energy electrons emits synchrotron radiation as well as IC
\grs very efficiently compared with the production rate of $\pi^0$-decay
\grs per CR nucleus and the overall IC emission can be of the same order
of magnitude as the hadronic \gr emission from SNRs \cite{mas}. In
fact, nonthermal electron synchrotron emission has been inferred for a
number of SNRs not only at radio wavelengths but also in hard X-rays, and
in recent years several such sources have also been reported in TeV \grs.
It was therefore as simple as it was tempting to interpret them in terms
of IC emission alone.
\vspace{3mm}

\noindent
\textbf{Theoretical Estimates of \gr Emission} 
\vspace{1mm}

\noindent
From the point of
view of CR nucleon origin, the \gr emission from electrons in SNRs appears
as a curse rather than a blessing, since it requires a separation of
hadronic from leptonic \grs. On the other hand we may as well turn this
fact into an advantage. Estimates of the \gr production in SNRs are based
on diffusive shock acceleration theory. Although one of the best developed
theories in astrophysics, its application to the time-dependent situation
of an evolving point explosion in a large scale magnetic field is
difficult. In addition, the overall evolution is a highly nonlinear
dynamical problem due to the high acceleration efficiency and the
backreaction of the accelerated particles on the structure of the shock
(e.g. \cite{mal}). Thus several sub-processes, like the strength of
injection of suprathermal particles into the acceleration process, which
cannot be calculated very accurately require observational input. Here the
synchrotron channel is important since the proton and electron spectra
have the same form for relativistic energies (apart from the trivial
radiation losses). In this way also the unknown magnetic field strength is
constrained.

The estimated flux values also depend on the character of the progenitor
star's evolution such as the mass loss for massive stars, and the type of
Supernova explosion, and they depend on the overall Supernova energetics.
Purely astronomical parameters, like source distance and ambient gas
density are obviously important as well. They can only be determined by
comprehensive multi-wavelength observations. And although certain
parameter combinations are fixed by the given thermal X-ray luminosity and
overall SNR dynamics, the combined uncertainty in the estimated \gr flux
may reach an order of magnitude. Given the marginal instrumental
sensitivity, it is not surprising that there were few detections and many
non-detections in the past five years. The history of these observational
efforts is quite interesting.
\vspace{3mm}

\noindent
\textbf{Early Observations: Upper Limits} 
\vspace{1mm}

\noindent
Early observations
($\sim 10$~hours) of radio-bright SNRs, associated with a number of
so-called unidentified EGRET sources, were unsuccessful. This concerned
especially the well-known core collapse SNRs $\gamma$-Cygni and IC-443.
Both the Whipple \cite{buc} and HEGRA \cite{voe97} found only
upper limits above a few 100 GeV. Extrapolations from the \gr energies $>
100$~MeV of the EGRET observations to 1 TeV are not necessarily
appropriate since the EGRET fluxes might be rather contaminated by Pulsar
emissions in that detector's very large field of view, cutting off beyond
about 20 GeV. The a priori flux estimates for the $\pi^0$-decay \grs just
about bracket the experimental upper limits.
\vspace{3mm}

\noindent
\textbf{Deep Observations of Historical Supernovae} 
\vspace{1mm}

\noindent
After
significantly deeper observations ($\sim 100$~hours) detections were
reported by the CANGAROO collaboration \cite{tan98}, \cite{tan01}
for the Type Ia Supernova SN 1006 from the year 1006 AD, the X-ray
brightest SNR in the Southern Hemisphere \cite{koy}  and presumably
the result of the deflagration of an accreting White Dwarf in a low
density environment, as well as for the X-ray-detected southern SNR RX
J1713.7-3946 \cite{mur}, possibly a core collapse SN in the
neighborhood of some dense gas clouds \cite{sla}. The HEGRA
collaboration \cite{ah01a} announced the detection of Cassiopeia A (Cas
A), the youngest Galactic SN from around 1680 AD, and
presumably the result of the core collapse of a massive Wolf-Rayet
progenitor. Cas A is also the brightest nonthermal radio source in the
sky. HEGRA \cite{ah01b} also reported a very low upper limit -- at 3
percent of the flux from the Crab Nebula -- for Tycho's SNR from 1572 AD,
a Type Ia Supernova seen by Tycho Brahe (on whose planetary orbit
observations Johannes Kepler based his laws of planetary motions). The
total observation time for Cas A was 230 h, the deepest \gr measurement at
TeV energies made up to know. All four objects had been been detected
before in the radio continuum and in hard X-rays, where also the latter do
presumably contain a significant synchrotron contribution.

SN 1006 has been phenomenologically modeled by many authors as an IC source
due to the X-ray synchrotron electrons in the Cosmic Microwave Background
(see e.g. \cite{tan98}). This presupposes a rather small interstellar
magnetic field of about 4 $\mu$G in order to explain the high \gr flux, given
the synchrotron flux, and then no $\pi^0$-decay \gr flux is needed. In the
opposite case of a significantly larger magnetic field, a significant
hadronic component is required, with a shell-type morphology because of the
gas compression at the shock \cite{aha99}.

Based on new data, the CANGAROO collaboration has recently reconsidered
its initial IC interpretation of the \gr emission from SNR RX J1713.7-3946
\cite{eno}. From a comparison of the inferred IC spectrum and
expected forms of a hadronic spectrum with the measured \gr spectrum it
was argued that the \gr nature of the object should rather be hadronic.
This claim did not remain uncontested, based on multi-wavelength arguments
\cite{rei}, \cite{but}. In the absence of an overall theoretical
model for this poorly understood source, for which is not clear whether it
is due to a Type Ia Supernova or due to a core collapse, it is difficult
to see how such controversies can be resolved.

\begin{figure}[t]
\begin{center}
\includegraphics[width=0.9\textwidth]{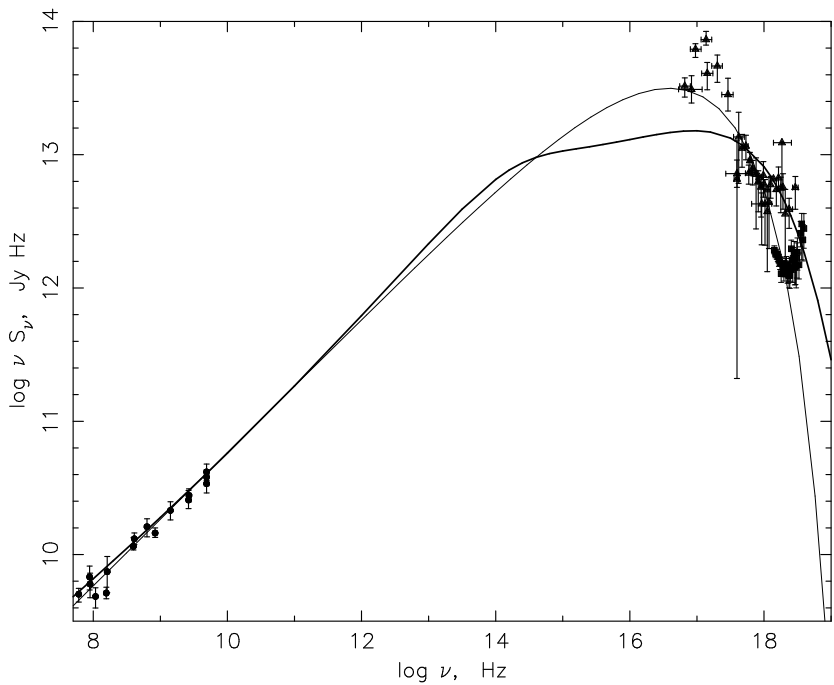}
\end{center}
\caption[]{Theoretical synchrotron spectral energy density for SN 1006
compared with radio and X-ray observations. The best fit is for efficient
proton acceleration, an effective upstream magnetic field of $20 \umu
$~Gauss and an electron to proton ratio $K_{ep}= 1.5 \times 10^{-3}$
(\emph{solid line}). The (physically not plausible) case of inefficient
proton
acceleration and low field strength of $4 \umu $~G (\emph{dashed line})
gives a
harder radio spectrum, high nonthermal X-ray emission and $K_{ep}= 4\times
10^{-2}$ (from \cite{ber02a})}
\label{ps4}
\end{figure}

\begin{figure}[t]
\begin{center}
\includegraphics[width=0.9\textwidth]{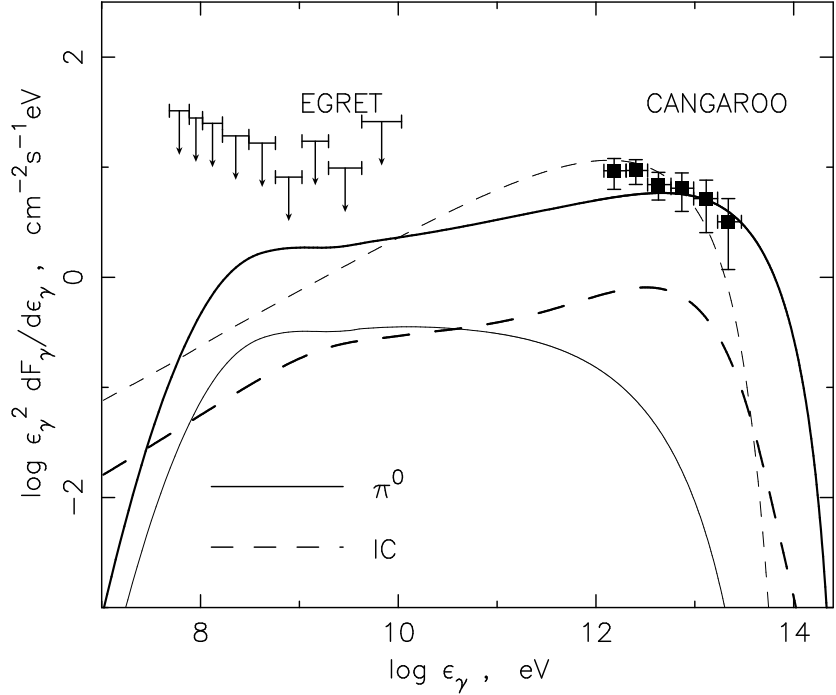}
\end{center}
\caption[]{Theoretical IC (\emph{dashed lines}) and $\pi^0$-decay
(\emph{solid
lines})
\gr spectral energy densities for SN 1006.  Efficient proton acceleration
is indicated by the thick curves, inefficient proton acceleration by the
thin curves. The recent high energy \gr flux data and the EGRET upper
limits \cite{nai} are also shown (from \cite{ber02a})} 
\label{ps5} 
\end{figure}

A self consistent estimate for SN 1006, based on time-dependent nonlinear
acceleration theory, calculates the space and time evolution of the
overall SNR dynamics together with the electron synchrotron, $\pi^0$-decay
and IC spectra, with the CMB as primary photon target \cite{ber99},
\cite{ber01a}, \cite{ber02a}. It shows (Fig.~\ref{ps4}) that not only a
high effective magnetic field strength of $\approx 20 \umu $G upstream of
the SNR shock is required to describe the overall synchrotron spectrum,
from the radio to X-rays \cite{rey}; \cite{ham}; \cite{all}, but that
one also needs a strong nonlinear shock modification due to efficient
acceleration of nuclei in order to explain the steep radio spectrum. This
is consistent with ion injection theory. The theoretical \gr spectrum is
then dominated by $\pi^0$-decay, even though only by a moderate factor of
order 5 (Fig.~\ref{ps5}), with the $\pi^0$-decay \gr spectrum extending up
to almost 100 TeV, and with a dipolar \gr emission morphology along the
external magnetic field direction. This predicted \gr spectrum agrees
reasonably well with the EGRET upper limits \cite{nai} and the latest TeV
results \cite{tan01}. with the $\pi^0$-decay \gr spectrum extending up to
almost 100 TeV.  An artificially assumed low magnetic field, combined with
a physically implausible low proton acceleration efficiency, describes the
synchrotron observations considerably poorer and the IC \gr spectrum
reaches less than about 10 TeV. As a consequence, good spatial as well as
spectral coverage from a minimum of 100 GeV and preferably even from about
100\, MeV up to the highest measurable \gr energies is needed to
ultimately resolve this issue for the high energy nuclear particles of
interest.

\noindent
It is also important that an analogous conclusion can be drawn from the
radio/X-ray spectra of Tycho's SNR which is calculated to lie just below
the present \gr detection limit \cite{voe02}. Also for Cas A -- a
more difficult object through its complex mass loss history before the
explosion -- the theory suggests a hadronic \gr emission
\cite{ber01b}. Recent work strongly supports this interpretation
\cite{ber02b}. It will be one of the important tasks for the coming
Northern Hemisphere Cherenkov telescopes to obtain good \gr spectra from
these sources.
\vspace{3mm}

\noindent
\textbf{HEGRA Galactic Plane Survey}
\vspace{1mm}

\begin{figure}[t]
\begin{center}
\includegraphics[width=0.35\textwidth,angle=-90]{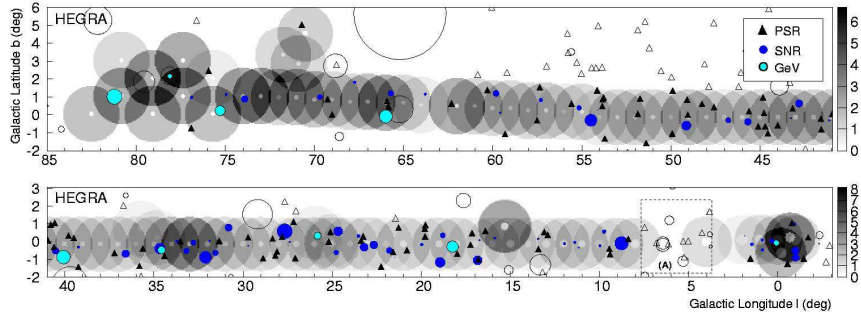}
\end{center}
\caption[]{Observation time in hours (\emph{right ordinate}) used for the
individual Galactic plane scan points given by the large \emph{gray
circles}.
\emph{Filled symbols} correspond to potential sources for which an upper
limit is
given. \emph{Symbol size} gives the size of the source. Objects in the
\emph{dashed
box}
were not included (from \cite{ah02a})}
\label{eps6}
\end{figure}

\noindent
A limited survey of the Galactic Plane (from longitude $85^{\circ}
 \mathrm{to} -3^{\circ}$) with the HEGRA system using on average 2.8
 hours of integration time per pointing (Fig.~\ref{eps6}) yielded
only upper limits for suspected individual sources.
The scan also included 19 known SNRs. Source stacking of the SNR
candidate sources gives a combined upper limit about a factor 2 above the
expected $\pi^0$-decay flux \cite{ah02a}. This shows that the
nondetections are
consistent with a dominant hadronic \gr emission.
\vspace{3mm}

\noindent
\textbf{Conclusions Regarding a SNR Origin for the Galactic Cosmic
Rays} 
\vspace{1mm}

\noindent
The search -- under the SNR lamp post -- for Galactic CR origin,
one of the problems of the century, has made remarkable progress through
high energy \gr astronomy during the last years. The few detections and
the many nondetections are consistent with a SNR origin of the dominant
nuclear CR component. We can expect that the new ground-based arrays,
coming on line these years, will decide this question.

\subsection*{Blazar Jet Emission}
\begin{figure}[t]
\begin{center}
\vspace{-2mm}
\includegraphics[width=.35\textwidth]{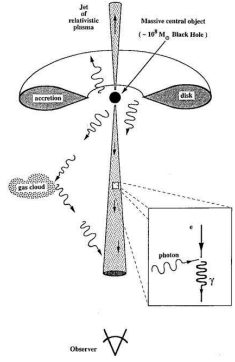}
\end{center}
\vspace{-4mm}
\caption[]{Schematic of a Blazar. An IC mechanism for the radiation from
the bulk relativistic AGN jet is assumed for definiteness, where the
ambient photon field may come from various sources, including internal
synchrotron photons produced in the jet itself (SSC). For Blazars
the jet is pointing close to the line of sight}
\label{ps7}
\end{figure}

Active Galactic Nuclei (AGNs) are presumably accretion-powered
supermassive Black Holes in the center of galaxies. The associated
jet-like outflows are often characterized by apparent superluminal
motion. Amongst the different AGNs there is a class of objects, the
Blazars, which in the optical show a dominant nonthermal continuum
together with broad lines, while being highly variable in time. The
continuum may be attributed to acceleration processes in the jet
(Fig.~\ref{ps7}).

Rather unexpectedly about 70 such objects have been found in the GeV
region by the EGRET instrument and this was one of the major discoveries
of CGRO. The high \gr luminosity that dominates the overall spectral
energy density distribution from the AGN suggests that the \gr emission is
strongly Doppler--boosted by coming from a relativistic jet
(\cite{bla}) whose bulk motion is essentially directed towards the
observer. In fact, the apparent luminosity $L_{\mathrm{app}}$ is
connected with the intrinsic source luminosity $L_{\mathrm{int}}$ through
$L_{\mathrm{app}} = \delta^4 L_{\mathrm{int}}$, where $\delta =
\Gamma^{-1}(1-\beta\, \mathrm{cos}\Theta)^{-1} > 1$ is the so-called
Doppler factor for bulk motion with Lorentz factor $\Gamma$ at an angle
$\Theta$ relative to the line of sight. In the limiting case of $\Theta$
going to zero $\delta = 2\Gamma$; $\delta$ might well be $\sim 10$.

Due to its small detection area the EGRET instrument had long integration
times which prevent the analysis of correlations with the fluxes in other
wavelength ranges on short time scales. Yet these are especially
interesting since the radiative electron cooling times at very high
energies are short enough so that their radiation amplitude may follow the
dynamical time variations of the system.

In a subclass of Blazars the optical lines are negligible or even absent.
These BL Lac objects, named after the prototype galaxy, show the maximum
of their synchrotron emission at X-ray energies. The corresponding \gr
emission is then expected to peak in the TeV region.

Several such objects have indeed been found in the TeV range. And since
the effective areas of the ground-based Cherenkov telescopes are very
large, they can follow rapid time variations much more effectively than a
space detector limited by photon statistics. Due to intergalactic
absorption TeV Blazars must be rather nearby at redshifts $z \ll 1$,
whereas many of the EGRET AGNs are distant, very luminous Quasars with a
flat radio spectrum. It is therefore perhaps not surprising that the
number of known TeV Blazars is a factor of 10 smaller. At distances of the
order of 150\,Mpc, corresponding to the well-measured objects Mkn 421 and
Mkn 501 at $z= 0.031$ and $z= 0.034$, respectively, the nonthermal
efficiency of such sources can be rather low. In order to avoid intrinsic
TeV \gr absorption in the radiating jet, $\delta$ should typically be of
order 10. Requiring the flux to be comparable to the Crab flux for
detectability, the intrinsic luminosity has to be roughly equal to
$L_{\gamma, \mathrm{intr}}^{\mathrm{source}} \sim (\delta /10)^{-4} \times
10^{40}\, \mathrm{erg/sec} \sim 10^{-5} \times L_{\mathrm{Edd}}
(10^7\,M_{\odot})$, where $ L_{\mathrm{Edd}} (10^7\,M_{\odot})$ is the
Eddington luminosity of a $10^7 M_{\odot}$ accreting object. The observed
variations are fast, down to sub-hour scales that translate to sub-parsec
spatial scales of the jet.

The question is then as to the nature of the jet emission. This is first
of all an interesting question in itself. Ultimately however, it should
also reveal the origin and composition of AGN jets. Presently the \gr
studies concern the nature of the jet emission and intergalactic
absorption.
\vspace{3mm}

\noindent
\textbf{Nature of the Jet Emission}
\vspace{1mm}

\noindent
The two main sources of radiation from jets can be energetic electrons
producing IC emission from low energy photon fields, or extremely high
energy protons that generate photo-pion and photo-pair cascades or
directly radiate synchrotron emission. 

\begin{figure}[t]
\begin{center}
\vspace{-10mm}
\includegraphics[width=.75\textwidth]{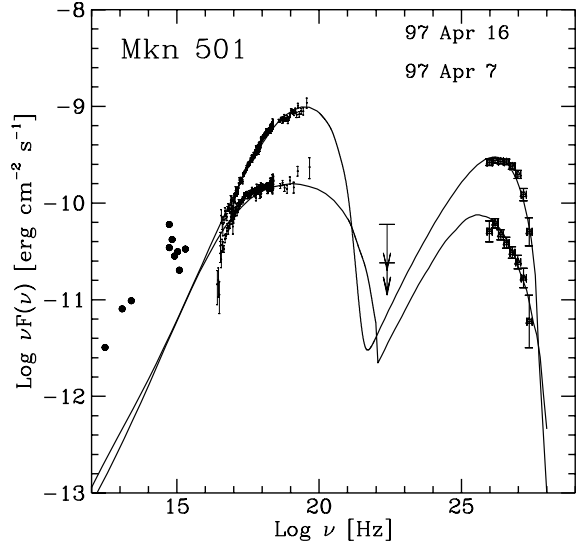}
\end{center}
\vspace{-8mm}
\caption[]{Multi-wavelength SSC modeling of the X-ray and TeV \gr energy
fluxes from Mkn\, 501, observed at different days with the {\it BeppoSAX}
satellite and the CAT telescope, respectively. Upper limits in the 100 MeV
region are from EGRET. The Inverse Compton peak lies in the Klein-Nishina
domain, with Intergalactic absorption being neglected (from \cite{tav})}
\label{ps8}
\end{figure}

At the comparatively low luminosities of BL Lac sources, the most
plausible IC target fields for leptonic jets are the radio synchrotron
photons from the same population of energetic electrons. The double peaked
spectrum typically inferred from quasi-simultaneous X-ray and high energy
\gr observations (e.g. \cite{tav}; Fig.~\ref{ps8}) then suggests
electron cooling to be about equally distributed between synchrotron and
IC losses and fast enough to explain the good temporal correlation
observed between the \gr with the X-ray fluxes. In such a synchrotron
self-Compton (SSC) interpretation the \gr flux should increase
quadratically with the synchrotron flux. This is about what is observed
(see also \cite{kra}).

Of course this interpretation is not unique. Alternative hadronic models
(e.g \cite{man}) require protons of extremely high energies $\leq
10^{19}$~eV in the jet. They produce pions on the abundant low frequency
photon fields longward of far infrared wavelengths, i.e. ultimately a
gamma signal and a extremely high energy neutrino signal. Such protons
could possibly also be the sources of the ultra-high energy CRs in our
Milky Way. And they may possibly power the high energy emission of flat
spectrum Quasars that appear optically thick for TeV emission.
Unfortunately the cooling of protons on photons is a rather slow process.
And for TeV \gr sources also a jet optical depth smaller than unity for
pair creation is required at TeV energies in order to allow the escape of
the photons produced in the interior. This limits the photon density
available for proton cooling.

Pion production has to compete with \emph{proton} synchrotron emission
which becomes important at such high energies \cite{ah2}. It
is a faster process for BL Lac objects as long as the magnetic field
strength in the \gr emission region is of order 100\, G. Then also
essentially all energy goes into TeV\, \grs. As a consequence, at least in
the TeV range, hadronic jets favor the proton synchrotron channel. Such
very large magnetic fields O(100)\, G are actually required for the
necessary fast acceleration, given the observed fast time variations. Such
high B-fields presumably require a massive cold hadronic jet component to
ensure dynamical equilibrium.
\vspace{3mm}

\noindent
\textbf{What are the jets made of?}
\vspace{1mm}

\noindent
It is quite surprising that
such an extreme alternative has not been resolved until now. Detailed
\emph{time-dependent} modeling of simultaneous multi-wavelength
observations will be necessary to clearly distinguish between leptonic and
hadronic jets in BL Lac objects. Obviously this is at the same time one of
the prerequisites for an understanding of the jet origin in the first
place.

\subsection*{Intergalactic \gr Absorption\\ 
on the Extragalactic Background Light}
The spectrum of the diffuse Extragalactic radiation field has a double
peak structure due to the direct radiation from stars and AGNs in the
UV, optical and near infrared on the one hand, and the mid and far
infrared reradiation of absorbed starlight by dust at longer wavelengths,
all integrated over the evolution of the Universe.

In this Extragalactic Background Light (EBL) TeV \grs are absorbed by pair
production with a cross section that peaks at about one quarter of the
Thompson cross section $\sigma_{\mathrm{T}}$. Therefore one can
approximately relate the two photon energies in the form $(E/1
\mathrm{TeV}) \approx (h \nu/1 \mathrm{eV})^{-1}$ and write the optical
depth in the form $\tau (E) = \xi (\sigma_\mathrm{T}/4)\, h\nu\, n_{
\mathrm{ph}}(h\nu) \times$\,distance, where $n_{\mathrm{ph}}(h\nu)$ is
the differential number density of the low energy photons and $\xi$ is of
order unity. Thus, for a constant spectral energy density of the EBL,
$\tau (E)$ increases linearly with $E$ and the TeV spectra from
Extragalactic sources will have an imprint in the form of characteristic
absorption features with a high energy cutoff. These absorption features
should give information on the spectrum of the EBL in an elegant way, an
information that direct observations of the EBL can only yield through a
difficult and uncertain subtraction of dominant foreground radiations such
as the Zodiacal Light or the so-called Galactic Cirrus; this subtraction
is especially problematic in the mid infrared region.

The uncertainty in this method lies in the unknown primary \gr source
spectrum. Thus for example Mkn\, 501 shows an exponential cutoff
proportional to
$E^{-1.9}\, \mathrm{exp}^{-E/E_0}$, with $E_0 = 6.2$~TeV which is roughly
consistent with observational estimates of the EBL spectrum
\cite{ah99}. However, on a $3 \sigma$ level, the source Mkn\, 421 (at
about the same redshift) appears to have a significantly lower cutoff
energy of $E = 3.6$~TeV \cite{ah02c}, precluding the possibility of
the cutoff being only an absorption feature of the EBL.

\begin{figure}
\begin{center}
\includegraphics[width=0.9\textwidth]{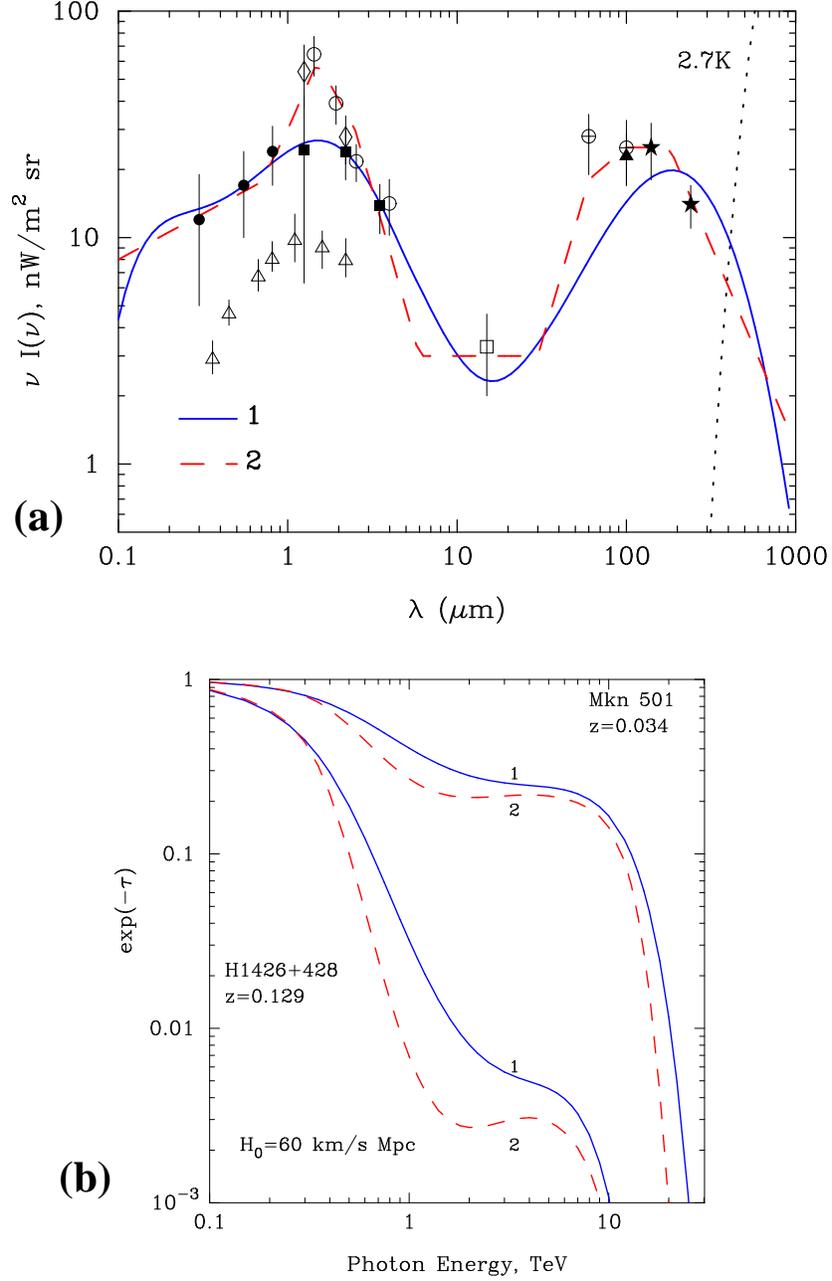}
\end{center}
\caption[]{Different empirical approximations to the direct measurements
of the spectral energy density of the EBL (\textbf{a}), and their
energy-dependent absorption effect on TeV photons from Mkn 501 and
H\,1426 + 428, respectively (\textbf{b}). The optical depth is denoted by
$\tau$ (from \cite{ah02c})}
\label{eps9ab}
\end{figure}

\begin{figure}
\begin{flushright}
\includegraphics[width=\textwidth]{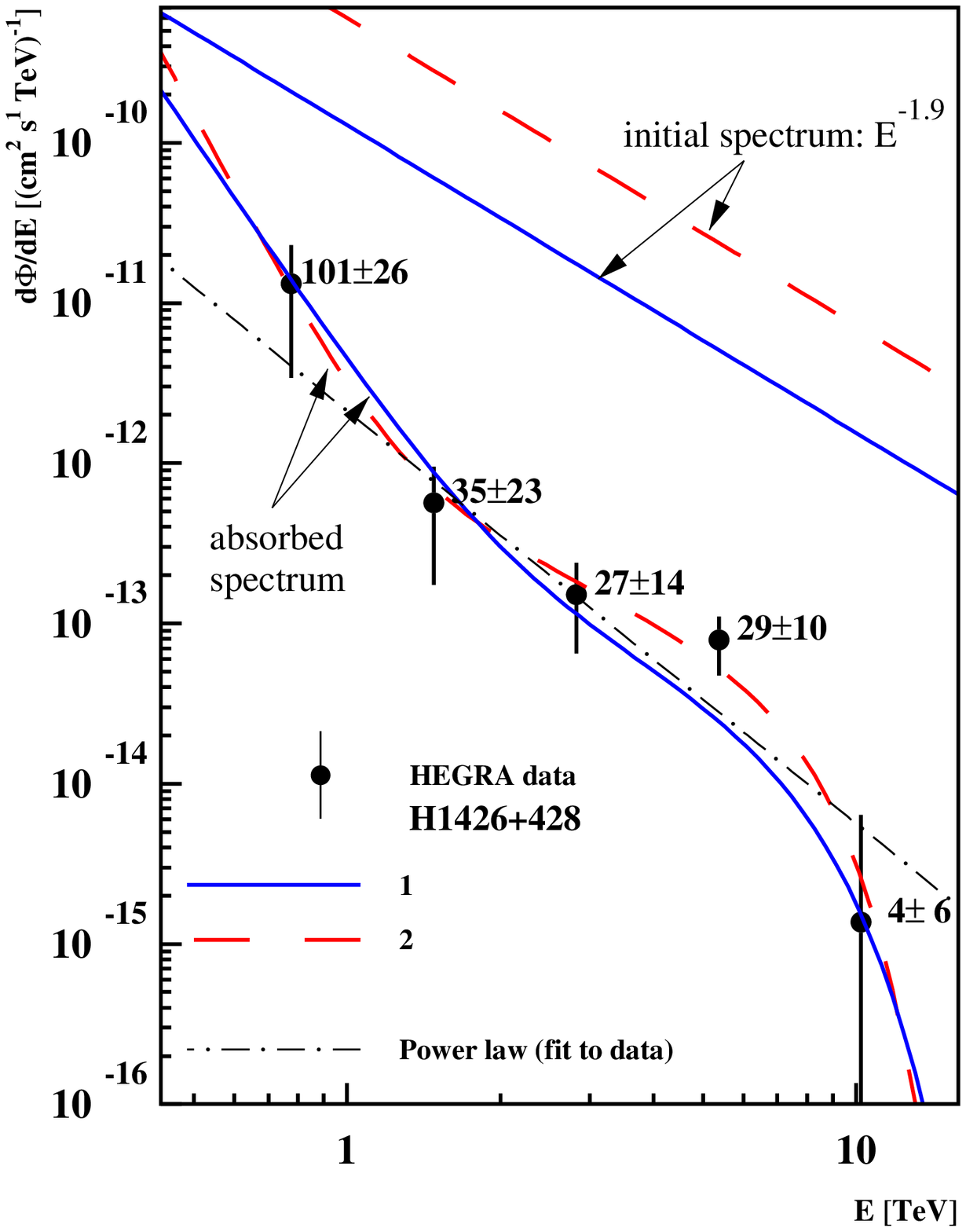}
\end{flushright}
\caption[]{Differential HEGRA spectrum of H\,1426 + 428 and its
approximation (\emph{solid} and \emph{dashed} curve) by the absorption
effect, cf.\ Fig.~\ref{eps9ab}, on initial (primary source) spectra
$\propto E^{-1.9}$ suggested by X-ray measurements. A power law fit is
given by the \emph{dash-dotted} curve (from \cite{ah02c})}
\label{eps10}
\end{figure}

Therefore we need to measure \gr sources at different, in fact higher
distances. Fortunately, the BL Lac object H\, 1426 + 428 at the fourfold
redshift $z = 0.129$ could recently be detected and its TeV spectrum
measured by HEGRA is within the errors {\it consistent with the
characteristic absorption features} expected \cite{ah02c}. The source
spectrum was assumed to be $\propto E^{-1.92}$, consistent with X-ray
synchrotron observations (Fig.~\ref{eps10}). The absorption feature
consists in a strong hardening of the observed \gr spectrum between about
2 and 5\, TeV (Fig.~\ref{eps9ab}).

%

One can also fit a power law to the HEGRA data alone -- with no further
justification than the simplicity of a 2-parameter straight line --
obtaining a somewhat lower overall statistical significance. However, this
power law is quite flat and it deviates therefore strongly from a steep
absorbed spectrum at energies below 1\,TeV. In fact, the Wipple
\cite{pet} and CAT \cite{dja02} telescopes have recently
confirmed the expected steep spectrum below 1 TeV.
\vspace{3mm}

\noindent
\textbf{Perspectives for the EBL from \gr measurements}
\vspace{1mm}

\noindent
These
results show clearly that TeV cutoffs alone contain insufficient
information, in contrast to earlier expectations \cite{ste}, because
cutoffs can also be mimicked by several effects, foremost by an intrinsic
cutoff of the source spectrum due to a finite maximum particle energy, or by
the Klein-Nishina effect. On the other hand the characteristic wavelength
variation of the absorption characteristics of the \gr spectra, measured at
different redshifts, offers the prospect of making accurate and convincing
\gr determinations of the EBL in the near and mid infrared in the near
future.

\section{Future Perspectives of High Energy \gr Astronomy}

\noindent
\textbf{Physics Questions}
\vspace{1mm}

\noindent
With the next generation of instruments coming
on line a much larger
number of sources will be detected. Such an increase by an order of
magnitude gives good reasons to expect that several of the major physics
problems which we have discussed here, will be solved. This should become
especially true, in one way or the other, for the origin of the Galactic
Cosmic Rays from SNRs. Another area of research, not mentioned above at
all, will be the 3-dimensional nonthermal structure of the Galaxy,
together with its halo in the form of the Galactic Wind (e.g.
\cite{ptu}). It should find its complement in investigations at low
radio frequencies with the proposed Square Kilometer Array.

Besides these developments \gr astronomy will increasingly move to
Extragalactic sources and to observational cosmology and, at least in
a serendipitous form, to Astroparticle Physics.

Beyond intrinsic AGN physics, the instrumental sensitivity increase will
allow studies of nearby starburst galaxies and through them the expected
formation of a strong nonthermal component \emph {throughout the Universe}
can be studied. Complementary studies aim at the nonthermal component of
galaxy clusters. By its large size and expected turbulent agitation, the
Intracluster Medium should not only confine the visible thermal matter and
the Dark Matter, but also the relativistic hadronic component since its
formation. This means that clusters of galaxies are closed systems,
preserving not only the chemical but also the nonthermal history and
entropy production since structure formation started \cite{vab}.

Strong emitters of very high energy \grs like the jets from flat spectrum
Quasars are expected to be surrounded by a halo of
$\E ^{+}\E ^{-}$~pairs due to the absorption of very high energy
\grs with $E_{\gamma} \sim 100$~TeV in the EBL and subsequent magnetic
isotropization in an intergalactic field $>10^{-12}$~G. The Compton
upscattering of photons from the Cosmic Microwave Background initiates a
cascade that becomes observable at lower \gr energies when the space
between us and the source becomes ultimately transparent \cite{ah94}.
The halos would be visible even if the jet points in an arbitrary
direction relative to the observer due to the magnetic isotropization.
Measurements of the angular size $\sim 1^{\circ}$ and \gr energy spectrum
of such halos should allow the determination of the Hubble constant, i.e.
the {\it absolute distance} of these objects, and a determination of the
{\it local} (in redshift $z$) EBL. Even though such measurements promise
to be difficult, and although there exists a substantial confusion problem
at larger redshifts, the possible rewards are correspondingly high.

It is also worth to emphasize the perspectives for Astroparticle Physics,
for instance by \gr observations of the Galactic Center region. One set of
simulations of the mass density $\rho(r)$ of Cold Dark Matter particles in
the gravitational potential well of the Galaxy suggests a rise $\varrho(r)
\propto r^{-1}$ with decreasing radius in the innermost region
\cite{nav}. However, other simulations come to more extreme results
(see e.g. \cite{kly} for a recent convergence study): $\varrho(r)
\propto r^{-3/2}$ for very small $r$, while agreeing for larger radii with
the $r^{-1}$-dependence. Depending on whether there is a strong density
cusp or not, the annihilation rate of e.g. Dark Matter neutralinos
$\propto \varrho^2$ in the very Galactic Center could therefore be quite
high or rather small (e.g. \cite{bucj}) and \gr observations of the
Galactic Center will not only be interesting from an astronomical point of
view. Calculations of the neutralino annihilation flux (e.g.
\cite{gon}) suggest the appearance of a line, possibly at energies
between 100\,GeV and 1\,TeV, besides a continuum that is strongly falling
off with \gr energy.  Fluxes may be at the percent level of the Crab
Nebula. Observations with the coming generation of Cherenkov telescopes
could thus provide a test of different halo models of our Galaxy and/or
put meaningful constraints on SUSY parameter space.
\vspace{3mm}

\noindent
\textbf{Next Generation Instruments}
\vspace{1mm}

\noindent
The next space projects in high energy \gr astronomy will be NASA's
Gamma-ray Large Astronomical Space Telescope
GLAST\footnote{\tt http://glast.gsfc.nasa.gov/.} with an expected launch in
2007, and its small brother, the Italian precursor mission Astrorivelatore
Gamma ad Immagini LEggero 
AGILE\footnote{\tt http://agile.mi.iasf.cnr.it/Homepage/.}, whose launch is
presently foreseen for 2004. Both detectors are based on silicon strip
technology. Comparable in sensitivity to EGRET, AGILE will have a much
larger FoV of 3\,sr and thus be very good for surveys. Similarly, it is
largely its survey capability which will distinguish GLAST from ground
based Cherenkov telescopes. However, GLAST will also be more than an order
of magnitude step beyond EGRET in sensitivity, angular resolution, and
spectral coverage. The energy range will extend to hundreds of GeV, even
though for reasons of statistics its de facto energy range will usually be
limited to some tens of GeV. GLAST will, first of all, be used to
investigate the large number of unidentified EGRET sources, left over from
the CGRO mission. Beyond, it is expected to find hundreds of AGNs, to
localize a fair number of SNRs, and to search for extended Extragalactic
objects.

\begin{figure}
\begin{center}
\includegraphics[width=\textwidth]{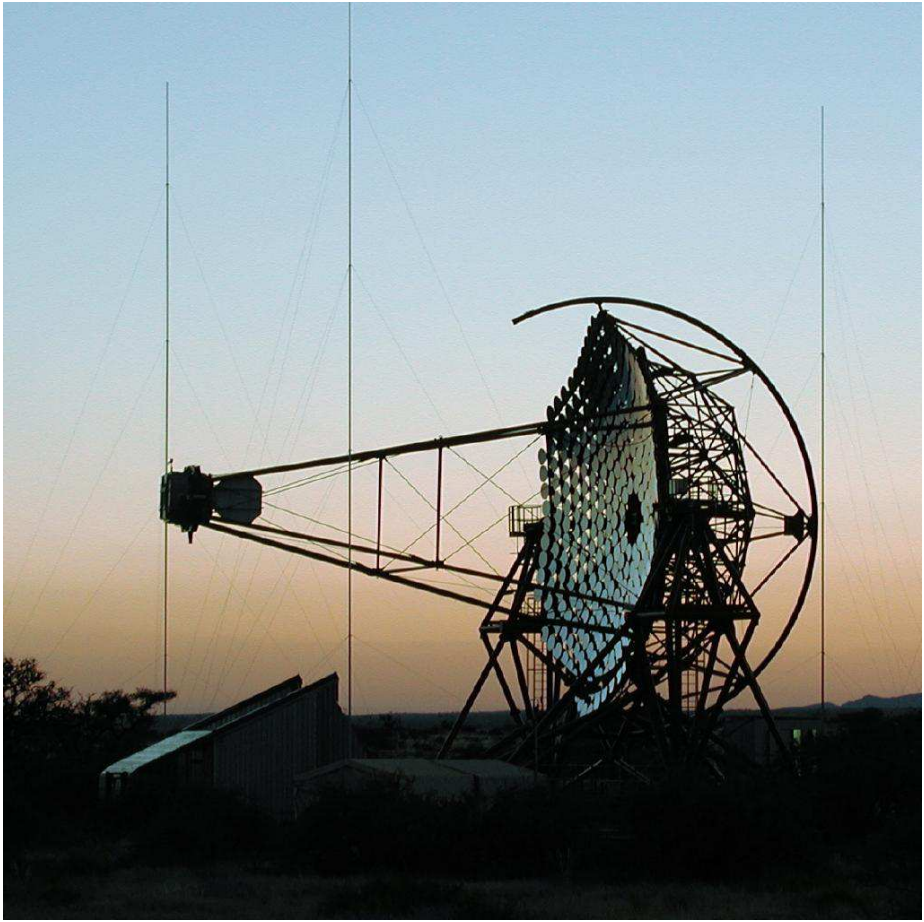}
\end{center}
\caption[]{The first of the four 12\, m telescopes of H.E.S.S. Phase I in
Namibia. The tessellated mirror consists of 380\, aluminized glass mirrors
of 60\,cm diameter. The focal plane detector (`camera') has 960\,
ultrafast photomultiplier pixels, covering an area of 1.4\,m that
corresponds to a field of view of $5^{\circ}$. The energy threshold is
about 100\,GeV and the sensitivity is about $10^{-12}$~erg/($\mathrm{cm}^2
\mathrm{s}$) above 100\,GeV and about $10^{-13}$~erg/($\mathrm{cm}^2
\mathrm{s}$) above 1
TeV for 50\,h of observation. (Photograph F. Toussenel, June 2002)}
\label{eps11} 
\end{figure}

On the ground, in the complementary energy region above $\sim 50$~GeV, the
future has already begun in Australia with the first 10\,m Cherenkov
telescope of the 2x2 stereoscopic array CANGAROO
III\footnote{\tt http://icrhp9.icrr.u-tokyo.ac.jp/c-ii.html.} operating since
more than one year. The next telescopes are expected to start observations
soon. In Namibia, the first 12\,m - telescope of the 2x2 array of the
Phase I of the High Energy Stereoscopic System
H.E.S.S.\footnote{\tt http://www.mpi-hd.mpg.de/hfm/HESS/HESS.html.} became
operational this June 2002 (Fig.~\ref{eps11}), to be followed by the other
three components in time steps of 6 months. The 17\,m single MAGIC
telescope\footnote{\tt http://hegra1.mppmu.mpg.de/MAGICWeb/.} is due to be
commissioned in La Palma still in 2002. And in a few years the 7-telescope
10\,m array
VERITAS\footnote{\tt http://veritas.sao.arizona.edu/veritas/index/shtml.} will
follow in Arizona.

Typically these instruments will have an energy threshold around 100\,GeV,
and an order of magnitude increase in sensitivity at 1\,TeV compared to
the previous generation instruments with their excellent angular
resolution of $0.1^{\circ}$. The energy resolution $\Delta E/E$ is about
15\,\%. Several of the ground based instruments will have made
detailed observations already years before GLAST. Nevertheless the
superior survey capability and the lower energy range will still reserve
GLAST important and unique goals.

The attraction of \gr astronomy at high energies is that it is a young
field. Whereas satellite instruments appear limited in their capabilities
simply by the required sizes and masses, this is not really true for
ground-based Cherenkov telescopes. Putting them on high mountain altitude,
like ESO's ALMA site at 5000\,m a.s.l., a future large extension in
threshold down to about 5\,GeV is possible with a 2x2 array of 20\,m
telescopes, while basically retaining the enormous effective area in the
$10^5~\mathrm{m}^2$ range \cite {akvq}. As a consequence, close-by \gr
bright objects like the Vela Pulsar could be detected in seconds with such
an array.


\section*{Acknowledgments}
I thank F.A. Aharonian, E.G. Berezhko, D. Breitschwerdt, L. Costamante, W. 
Hofmann, D. Horns, and M. Panter for valuable discussions, and J. 
Suppanz-Pirsch for expert help with the manuscript

%

\end{document}